\documentclass[sigchi-a,nonacm]{acmart}
\usepackage{framed}
\usepackage{enumitem}
\AtBeginDocument{%
  \providecommand\BibTeX{{%
    \normalfont B\kern-0.5em{\scshape i\kern-0.25em b}\kern-0.8em\TeX}}}

\begin{document}

%%
%% The "title" command has an optional parameter,
%% allowing the author to define a "short title" to be used in page headers.
\title{Why should we care about register? Reflections on chatbot language design}

%%
%% The "author" command and its associated commands are used to define
%% the authors and their affiliations.
%% Of note is the shared affiliation of the first two authors, and the
%% "authornote" and "authornotemark" commands
%% used to denote shared contribution to the research.
\author{Ana Paula Chaves}
%\authornote{Both authors contributed equally to this research.}
\email{anachaves@utfpr.edu.br}
\orcid{0000-0002-2307-3099}
\affiliation{%
  \institution{Federal University of Technology--Parana}
  \streetaddress{R. Rosalina Maria Ferreira, 1233}
  \city{Campo Mourao}
  \state{Parana}
  \country{Brazil}
}

\author{Marco Aurelio Gerosa}
\affiliation{%
  \institution{Northern Arizona University}
  \streetaddress{1295 Knoles Dr}
  \city{Flagstaff}
  \state{AZ}
  \country{USA}}
\email{Marco.Gerosa@nau.edu}

%%
%% By default, the full list of authors will be used in the page
%% headers. Often, this list is too long, and will overlap
%% other information printed in the page headers. This command allows
%% the author to define a more concise list
%% of authors' names for this purpose.
\renewcommand{\shortauthors}{Chaves and Gerosa, et al.}

%%
%% The abstract is a short summary of the work to be presented in the
%% article.
\begin{abstract}
This position paper discusses the relevance of register as a theoretical framework for chatbot language design. We present the concept of register and discuss how using register-specific language influence the user's perceptions of the interaction with chatbots. Additionally, we point several research opportunities that are important to pursue to establish register as a foundation for advancing chatbot's communication skills.
\end{abstract}

%%
%% The code below is generated by the tool at http://dl.acm.org/ccs.cfm.
%% Please copy and paste the code instead of the example below.
%%


%%
%% Keywords. The author(s) should pick words that accurately describe
%% the work being presented. Separate the keywords with commas.

%%
%% This command processes the author and affiliation and title
%% information and builds the first part of the formatted document.
\maketitle

\section{Introduction}

In task-oriented interactions, chatbots often assume social roles traditionally associated with a human service provider, for example, tutor~\cite{tegos2016investigation}, salesperson~\cite{gnewuch2017towards}, and tourist assistant~\cite{chaves2018single}. When a chatbot uses unexpected levels of (in)formality or incoherent language style, the interaction may result in frustration or awkwardness~\cite{kirakowski2009establishing, elsholz2019exploring}. To convey competence and be recognized by human-interlocutors for the role they stand in, chatbots' language must cohere with the situation in which the interaction takes places and the social role they aim to represent. To this date, there is no formal techniques to guide the design of a chatbot's language, which is often based on the designer's personal linguistic habits or an ad-hoc analyses of personas and user characteristics.

\begin{sidebar}
    \begin{framed}
        \textbf{Register according to~\citet{biber1988variation}:} the language variety associated with a particular situation of use.\vspace{0.75pc}\\
        \textbf{Register in this paper:} the \textit{core linguistic features} in a conversation, given the \textit{context}, where the linguistic features consist of the set of words or grammatical characteristics that occur in the conversation, and the context consists of a set of \textit{situational parameters} that characterize the situation in which the conversation occurs, e.g., the participants, the channel, the production circumstances, and so on.
    \end{framed}
    \caption{Definition of Register}
    \label{margin:register}
\end{sidebar}

Language can often be accounted for by factors such as style (e.g.,~\cite{leech2007style}), dialect (e.g.~\cite{szmrecsanyi2011corpus}), genre (e.g.,~\cite{kamberelis1995genre}), and register (e.g.,~\cite{conrad2009register}). Style is by far the factor that receives more attention from researchers on conversational agents~\cite{feine2019taxonomy, thomas2018style, jakic2017impact, lin2017stylistic, syed2020adapting}. In sociolinguistics, style is defined as the linguistic variation that reflects \textit{aesthetic} preferences, usually associated with particular speakers or historical periods \cite{conrad2009register} (e.g., archaic vs. modern English). However, studies emphasize that the \textit{``core linguistic features, like pronouns and verbs, are \textit{functional}''} rather than aesthetic~\cite{conrad2009register}, which points to the concept of \textbf{register}.

\begin{marginfigure}
    \centering
    \includegraphics[width=0.4\textwidth]{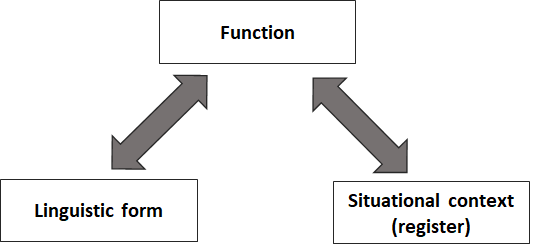}
    \caption{Representation of register}
    \label{fig:register}
\end{marginfigure}

According to the register theory, language variation is situationally-defined, meaning that every utterance in a conversation is influenced by situational parameters, such as the relationship between participants, the purpose of the interaction, and the topic of the conversation~\cite{kamberelis1995genre, conrad2009register}. Register theory aims to link the occurrences of certain linguistic features in utterances to the situational parameters of the conversation~\cite{conrad2009register}. Figure~\ref{fig:register} illustrates the relationship among language variation, register, and function~\citep{egbert2016all}.

Register is an established theory in the sociolinguistics domain (see e.g., \citep{biber1988variation, biber2012register}), and has been shown as a reliable predictor of language variation across conversational contexts~\citep{biber2012register}. Despite its importance for understanding language variation in human-human communication, it has not been widely explored in the context of human-chatbot interactions. In this position paper, we argue on the applicability of register theory as a theoretical foundation for tailoring chatbot linguistic choices to a particular interactional situation.

\section{Register for chatbot language design}

A variety of factors influence how people perceive a chatbot's social skills~\citep{chaves2020how} and, as user expectations of proficiency increase, one important way to enhance chatbot interactions is by carefully planning their use of language~\citep{chaves2020how, kirakowski2009establishing, go2019humanizing}. User perceptions of chatbot conversational skills are important because chatbots are targeted to fluidly interact using natural language. Chatbots are often deployed to perform social roles traditionally associated with humans, particularly in contexts where there may be consequences for a human if they choose to act on the chatbot's information. This means that user perceptions of chatbot competence and credibility are crucial for a chatbot's success~\citep{zumstein2017chatbots}.

Previous studies have found that appropriate language style is not relevant for determining user satisfaction as long as the user can understand the chatbot's answer, only advising that the chatbot's language style should be ``mildly appropriate to the service the chatbot provides''~\citep{balaji2019assessing}. In a previous study~\citep{chaves2021chatbots}, however, we found evidence that user experience goes beyond merely comprehending a chatbot's utterance to whether the user perceives the chatbot's language as appropriate and credible. In that study, we explored the applicability of register to human-chatbot interactions by developing a rationale for accounting for register in chatbot design, and providing a mechanism for implementing theory into design practice. The results demonstrated that register characteristics are more relevant than individual preferences or personal habits for perceived appropriateness of language, credibility, and overall user experience. Thus, register characteristics can be seen as primary drivers for perceptions of language, and designers should consider register to foster chatbot acceptance and success.

Therefore, we argue that user perceptions are also shaped by \textit{how} information is conveyed--as characterized by the conversational register--and register theory can provide a sound theoretical framework for concretely characterizing the language use and analytically exploring how variations in the patterns of language impact user perceptions of the quality of a chatbot, including critical factors, like appropriateness and credibility.

\section{Implications for chatbot design}
\label{sec:implications}

In this position paper, we argue that register analysis should be used as a technique for providing a theoretical basis for chatbot language design, grounded on the results presented in~\citet{chaves2021chatbots}. To design register-specific language for chatbots, we need to enrich chatbots with computational models that can adapt the utterances to conform with the expected register, in an effort to mimic the subconscious humans' language production process. %This has important implications for designers of the next generation of chatbots.

When retrieving text from external sources, utterances should be adapted to the new conversational situation in which the chatbot is embedded. This is not generally done in the current generation of chatbots; it is common to find chatbots that extract and present information directly from websites, books, or manuals without any linguistic adaptation to the new context. For example, Golem\footnote{Available in Facebook Messenger at http://m.me/praguevisitor. Last accessed: June, 2020} is a chatbot designed to guide tourists through Prague (Czech Republic); its utterances are extracted from an online travel magazine\footnote{https://www.praguevisitor.eu} without any adaptation to the new interactional situation (which differs in production, channel, and setting).

New generations of chatbots will be expected to generate their own custom-constructed utterances dynamically, which will require sophisticated natural language engines that are able to adapt dynamically their conversational register to changing situational parameters. In this context, baseline corpora for training the conversation models will be required at the heart of such natural language engines. Recent studies~\citep{chaves2019chatting} showed that using corpora in the same domain is not enough to account for language appropriateness, as differences in the interlocutor's social role and communication purpose result in varying patterns of language~\citep{chaves2019chatting}. Designers should carefully ensure that the register found in a corpus used to train such models matches the register implied by the situational parameters, or they need to use algorithms to adapt the language accordingly.

Using register analysis to characterize different situations and how they map to the most appropriate register profiles can provides a way forward in the design of the next generation of chatbots; one could imagine a chatbot language engine that, given a particular situational profile for planned conversations, could automatically configure its language to present information in the most appropriate register. The work presented by ~\citet{chaves2021chatbots} is a first step toward the applicability of register theory in chatbot language design. In the next section, we point out some research opportunities.

\section{Research opportunities}

There are open challenges that need to be addressed in future work to formalize register as a theoretical basis for chatbot's language design:

\begin{description}[leftmargin=0cm]
    \item[Text modification automation:] when a chatbot retrieves information from external sources, it is necessary to adapt the utterance to the new situational parameters, which requires modifying the distribution of particular linguistic features in the sentence. This process can be achieve through text modification, which consists of changing the language of a text while preserving the text's content and integrity~\citep{sato2007guide, bosher2008effects}. Research on style transfer has made progress on reproducing the patterns of language in a new corpus (see e.g., \citep{syed2020adapting}; however, most algorithms assume parallel data with similar content distributions. To fully automate text modifications, we would need to develop algorithms that, given the frequencies of words expected in a register, perform text modifications in an utterance to approximate the frequency of words to the frequencies in the expected register. 
    \item[Extensibility of register characteristics to similar domains:] register characterization is situational-dependent. Identifying the register characteristics for every possible interaction context may be overwhelming. Thus, it is crucial to understand how register characteristics can be extended to different domains. We expect that, as we reduce the differences in situational characteristics, we minimize the language variation across registers. In this sense, one interesting future research includes performing situational analysis of several domains where chatbots are commonly used and find the intersections in the situation parameters. Then, one can empirically evaluate the extent to which the generalization to similar domains would apply.
    \item[Influence of sentence structure:] the study presented in~\citet{chaves2021chatbots} analyzed the influence of linguistic features individually without considering the position where they occur in the sentence or the frequent co-occurrences with other features. We are not aware of a methodology that automatically weighs the features according to their place in the sentence. Developing such methodologies would potentially improve the ability to detect the factors that influence user perceptions, which likely include sentence structure and flow.
    \item[Ranking sentences according to the register:] to allow machine generation of register-specific language, it is crucial that we develop statistical models that, given a set of candidate sentences for a particular situation, rank them according to their coherence with the expected register. This model would select the sentence that has the highest frequency of preferred features and lowest frequency of unfavored features as the top rank position. This is still unexplored in the current literature.
    \item[Corpus selection automation:] Section~\ref{sec:implications} pointed out the importance of finding a register-specific reference corpus for chatbot language generation. As a consequence, it is important to create a tool that allows practitioners to find the appropriate reference corpus when designing chatbot utterances. We suggest a search platform in which researchers and practitioners could publish corpora of conversations that could be used for particular situations as well as search for a reference corpus for designing their own chatbot. The search would consider the situational parameters in which the conversation takes place to suggest a corpus for training that complies with the expected register. Such platform would be relevant not only for the chatbot industry but also for researchers in several domains such as Natural Language Processing and Generation, Corpus Linguistics, and Computational Linguistics.
    \item[Fairness of language:] this paper suggests that chatbot language can be designed by training a chatbot using a corpus of register-appropriate conversations. Corpus-based language generation must, however, consider ethical concerns, such as the fairness of language. Scholars~\cite{schlesinger2018let, marino2014racial, marino2006chatbot} have pointed out the risks of using biased contents of databases to generate language. Unlike style, register focuses on the distribution of linguistic features associated with the context rather then personal linguistic habits. Nevertheless, future research could focus on identifying possible biases introduced by register-specific language generation in order to avoid negative implications of using biased language.
\end{description}

In conclusion, we expect that this position paper open new research avenues that bring together researchers in computer science, machine learning, and linguistics to design chatbots that use register-specific language and, consequently, provides an enriched user experience.
%%
%% The acknowledgments section is defined using the "acks" environment
%% (and NOT an unnumbered section). This ensures the proper
%% identification of the section in the article metadata, and the
%% consistent spelling of the heading.
%%
%% The next two lines define the bibliography style to be used, and
%% the bibliography file.
\bibliographystyle{ACM-Reference-Format}
\bibliography{doutorado}

%%
%% If your work has an appendix, this is the place to put it.
%\appendix

\end{document}